\documentstyle[prl,aps,multicol,epsfig]{revtex}
\begin{document}
\title{Coherently Scattering Atoms from an Excited Bose-Einstein Condensate}
\author{M.J. Bijlsma and H.T.C. Stoof \\
Institute for Theoretical Physics, University of Utrecht, \\
Princetonplein 5, 3584 CC Utrecht, The Netherlands}
\maketitle
\begin{abstract}
We consider scattering atoms from a fully Bose-Einstein condensed gas.
If we take these atoms to be identical to those in 
the Bose-Einstein condensate,
this scattering process 
is to a large extent analogous to Andreev reflection 
from the interface between a
superconducting and a normal metal.
We determine the scattering wave function 
both in the absence and the presence 
of a vortex. Our results show a qualitative
difference between these two cases that can be understood
as due to an Aharonov-Bohm effect. It leads to the possibility 
to experimentally detect and study vortices in this way.
\end{abstract}
\begin{multicols}{2}
The achievement of Bose-Einstein condensation in trapped alkali vapors
has generated renewed theoretical and experimental interest in the 
subject of quantized vortices, since
the quantization of circulation, as well as the presence of
persistent currents, are amongst the most striking features
of superfluidity in a Bose-condensed fluid. 
In the case of superfluid helium,
vortices have been studied extensively 
\cite{onsager,feynman,yarmchuk,packard}.
Because of the strongly interacting
nature of such a liquid however, 
the understanding of these phenomena is mainly phenomenological.
In contrast, in weakly-interacting Bose condensed gases 
the theoretical study of vortices can be based on a microscopic theory.

Starting from such a microscopic viewpoint, problems that have 
been studied in detail 
are the stability of a vortex in a harmonic trapping potential,
the related question of its experimental creation, 
and the detection of these vortices 
\cite{stringari99,baym96,dalfovo96,rokhsar97,fetter,dodd97,lundh,bolda,tempere,sinha,zambelli,svidzinsky,goldstein98,feder99,vortexpriv}. The   
creation of a stable vortex appears to be possible by
rotating an external trapping potential that is sufficiently anisotropic
with a frequency larger than some critical value 
\cite{stringari99,dalfovo96}.
In this case, a vortex configuration becomes the
thermodynamic equilibrium of the system.  
This equilibrium can be reached in two ways, either
by first cooling the gas through the 
transition temperature and then rotating the trapping potential, or 
by first rotating and then cooling the gas.
The latter seems to be the most favorable experimentally \cite{stringari99}. 
Concerning the detection of a vortex, 
several ideas have been put forward. 
It might be possible to detect the core of the vortex
after the Bose-condensed cloud is released from the trap
and has been allowed to expand for some time \cite{lundh}. 
Also, the interference
fringes between two condensates containing vortices
will exhibit dislocations \cite{bolda,tempere}. In addition, the excitation
frequencies of the collective modes of the condensate
are shifted due to the presence of a 
vortex \cite{sinha,zambelli,svidzinsky}. 
Finally, it has been proposed to study vortices by optical means
\cite{goldstein98}.

In addition to these ideas, in this paper we suggest and investigate a
method which is based on the scattering of identical 
atoms off the Bose-condensed cloud, 
analogous to the occurrence of Andreev reflection \cite{andreev}
from a superconductor-normal metal interface \cite{comment}.  
In the latter case, the scattered electrons probe both the magnitude 
and the phase of the superconducting order parameter.
A similar experiment with a trapped Bose condensate will, 
besides the condensate density, also probe 
gradients in the phase of the Bose-condensed cloud.
This is desirable, since the vortex core is typically  
small compared to the overall size of the condensate,
whereas the superfluid velocity due to the presence 
of the vortex extends over the entire
Bose condensed cloud. Because the velocity field of a condensate 
with and without a vortex is qualitatively different, 
the scattering of atoms might
be a possible way to detect a vortex. 

Note that a similar idea has been put forward
recently in the context of superfluid $^{4}$He, 
to study the quantum sticking, scattering and transmission
of $^{4}$He atoms from superfluid $^{4}$He slabs \cite{campbell98}.
In an experiment with a Bose-condensate, 
the energy of the incoming atoms
has to be rather small to see an effect, 
namely of the order of the chemical potential of the
condensate. A convenient way to realize this experimentally is
by using optical means to extract the atoms 
that are to be scattered from the condensate 
itself. After displacing these atoms with respect to the condensed
cloud, they are subsequently released and measured by
the usual time-of-flight measurements or {\it in situ} 
optical imaging  \cite{cornellpriv}.

Bearing this in mind, we consider the whole scattering process
to take place in an external confining potential. 
Such a process can be described by the 
time-dependent Bogoliubov-deGennes equations.
To lowest order approximation, 
the condensate density profile can be taken to be
static, which implies that the
time dependence of the mean-field interaction due to the
possible excitation of one or more collective modes is neglected.
In the static approximation, 
the coherence factors $f({\bf r},t)$  and $g({\bf r},t)$ 
for the atom satisfy

\begin{eqnarray}
\label{eq1}
i \hbar {\partial \over \partial t}
\left[ \!
\begin{array}{c}
f({\bf r},t) \\
g({\bf r},t) 
\end{array} 
\! \right] 
& \!=\! & 
\left[ \!
\begin{array}{cc}
H & V \\
- V^{*} & -H 
\end{array}
\! \right] \!
\left[ \!
\begin{array}{c}
f({\bf r},t) \\
g({\bf r},t)
\end{array} 
\! \right] \; ,
\end{eqnarray}
where

\begin{mathletters}
\begin{eqnarray}
H & = &
-{\hbar \nabla^{2} \over 2 m} + V_{ext}({\bf r}) - 
\mu + 2 T^{2B} \rho({\bf r}) 
\end{eqnarray}
and
\begin{eqnarray}
V & = & T^{2B}  \rho({\bf r}) \exp\left[2i\chi({\bf r})\right] 
\; . 
\end{eqnarray}
\end{mathletters}
Here, $T^{2B}\!=\!4\pi\hbar^{2}a/m$
is the $s$-wave approximation for the two-body scattering matrix
and $V_{ext}({\bf r})=\sum_{i} m \omega_{i}^{2} x_{i}^{2}/ 2$ 
denotes the external trapping potential.
The inter-atomic scattering length is given by $a$.
Furthermore, the static density profile of the condensate is denoted by 
$\rho({\bf r})$ and its phase by $\chi({\bf r})$.
Outside the condensate, $g({\bf r},t)$ has to be zero and
Eq.~(\ref{eq1}) reduces to the Schr\"odinger equation for an atom 
in a harmonic potential.

Of course, the colliding atom can in principle 
excite the condensate.
To include the possibility of exciting collective modes,  
we simply have to replace $\rho({\bf r})$ and $\chi({\bf r})$
by their operator counterparts,
$\hat \rho({\bf r},t)$ and $\hat \chi({\bf r},t)$, respectively.
The latter two can be expanded in terms
of the creation and annihilation operators $\hat b_{i}^{\dagger}$
and $\hat b_{i}$ of the collective modes with  
mode functions $u_{i}({\bf r},t)$ and $v_{i}({\bf r},t)$ as   

\begin{mathletters}
\begin{eqnarray}
\hat \rho({\bf r},t) & \! = \! & \rho({\bf r}) \! + \! \sqrt{\rho({\bf r})}
\sum_{i} \left[ u_{i}({\bf r},t) \! - \! v_{i}({\bf r},t) \right]
 \left[ \hat b_{i} \!+\!
\hat b^{\dagger}_{i} \right] 
\end{eqnarray}
and
\begin{eqnarray}
\hat \chi({\bf r},t) & \! = \! & \chi({\bf r}) \! + \! 
{i \over 2 \sqrt{\rho({\bf r})}}
\sum_{i}  \left[ u_{i}({\bf r},t) \!+\! v_{i}({\bf r},t) \right]
\left[ \hat b_{i} \!-\! \hat b^{\dagger}_{i} \right] \; .
\end{eqnarray}
\end{mathletters}
If we denote by $|\alpha\rangle$
the number states in the Fock-space spanned 
by the creation and annihilation operators, 
the different channels of our 
scattering problem are given by 

\begin{eqnarray}
\left[
\begin{array}{c}
f^{\alpha}({\bf r},t) \\
g^{\alpha}({\bf r},t)
\end{array}
\right] \otimes |\alpha\rangle \; ,
\end{eqnarray}
and the equations of motion for these different channels 
read

\begin{eqnarray}
i \hbar {\partial \over \partial t}
\left[ \!
\begin{array}{c}
f^{\alpha}({\bf r},t) \\
g^{\alpha}({\bf r},t) 
\end{array} 
\! \right] 
& \!=\! & 
\sum_{\alpha'}
\langle \alpha| \left[ \!
\begin{array}{cc}
\hat H & \hat V \\
- \hat V^{\dagger} & -\hat H 
\end{array}
\! \right] \!
|\alpha' \rangle 
\left[ \!
\begin{array}{c}
f^{\alpha'}({\bf r},t) \\
g^{\alpha'}({\bf r},t)
\end{array} 
\! \right] \; ,
\end{eqnarray}
where 

\begin{mathletters}
\begin{eqnarray}
\hat H & = &
-{\hbar \nabla^{2} \over 2 m} + V_{ext}({\bf r}) - 
\mu + 2 T^{2B} \hat \rho({\bf r}) 
\end{eqnarray}
and
\begin{eqnarray}
\hat V & = & T^{2B}  \hat \rho({\bf r}) \exp\left[2i\hat 
\chi({\bf r})\right] \; . 
\end{eqnarray}
\end{mathletters}
Again, outside the condensate $g^{\alpha}({\bf r},t)$ has to be zero.
It is clear that if the condensate is initially in 
the ground state $|0\rangle$, after colliding with the incoming atom
it can be in the ground state or any of the excited states.
Note that $|0\rangle$ can in principle be an arbitrary
solution of the Gross-Pitaevskii 
equation and can in particular also be a vortex or a kink solution.

In the remainder of this work,
we apply the static approximation and include only the ground state
$|0\rangle$. The resulting equations describing our 
scattering problem then reduce to Eq.~(\ref{eq1}).
This is expected to be accurate in the case
of the scattering of a single atom at energies
several times larger than the chemical potential. At these
energies, the kinetic energy of an incoming particle
is still relatively large and the overlap of its
wavefunction with that of the low-lying collective modes small.
In addition, we take the external potential to be
rotationally symmetric around the z-axis, i.e.,
$\omega_{x}=\omega_{y} \equiv \omega_{r}$, and 
approximate the cigar-shaped traps used in experiments
by assuming translational invariance in the z-direction.
It is then convenient to write the Laplace operator in cylindrical 
coordinates $(r,z,\phi)$, i.e.,
$\nabla^{2} = {1 \over r} {\partial \over \partial r}
+ {\partial^{2} \over \partial r^{2}} + 
{1 \over r^{2}} {\partial^{2} \over \partial \phi^{2}} +
{\partial^{2} \over \partial z^{2}}$
and expand $f({\bf r},t)$ and $g({\bf r},t)$ as

\begin{mathletters}
\begin{eqnarray}
f(r,\phi,t) & = & 
\sum_{n} {f_{n}(r,t) \over \sqrt{r}} e^{i n \phi} 
\end{eqnarray}
and
\begin{eqnarray}
g(r,\phi,t) & = & \sum_{n} {g_{n}(r,t) \over \sqrt{r}} e^{i n \phi} \; .
\end{eqnarray}
\end{mathletters}
Inserting this expansion into Eq.~(\ref{eq1}),
the Bogoliubov-deGennes equations 
for the wavefunctions $f_{n}(r,t)$ and $g_{n}(r,t)$ become,

\begin{eqnarray}
i \hbar {\partial \over \partial t}
\left[ \!
\begin{array}{c}
f_{n} \\
g_{n}
\end{array} 
\! \right] 
& \!=\! & 
\sum_{n'}
\left[ \!
\begin{array}{cc}
H_{nn'} & V_{nn'}  \\
- V_{n'n} & - H_{nn'}
\end{array}
\! \right] \!
\left[ \!
\begin{array}{c}
f_{n'} \\
g_{n'}
\end{array} 
\! \right] \; .
\end{eqnarray}
Here, the matrix elements 
$H_{nn'}$ and $V_{nn'}$ equal  

\begin{mathletters}
\begin{eqnarray}
\label{eq6}
H_{nn'} 
& = & \left[ 
-{\hbar^{2} \over 2m} {\partial^{2} \over \partial r^{2}} 
+V_{ext}(r)- \mu \right. \\ 
& & \left. \; + {\hbar^{2}(n^{2}-1/4) \over 2 m r^{2}} +  
2 T^{2B} \rho(r)
\right] 
\delta_{n,n'} \nonumber \; , 
\end{eqnarray}
and
\begin{eqnarray}
V_{nn'} & = & T^{2B} \rho(r) \delta_{n,n'+2l} \; .
\end{eqnarray}
\end{mathletters}
Note that we consider the solution of Eq.~(\ref{eq1})
in the presence of a vortex with a winding number $l$
which is aligned in the direction of the z-axis.
This implies that the condensate wavefunction is given by
$\psi({\bf r},t)=\sqrt{\rho(r,z,t)} e^{il\phi}$.

Upon solving the resulting equations
we treat the condensate in the Thomas-Fermi approximation,
both in the presence and in the absence of a vortex.
Thus, the density profile and the
phase are given by

\begin{mathletters}
\begin{eqnarray}
\rho(r) & \!=\! & {\left[
\mu \!-\! V_{ext}(r) \!-\! {\hbar^{2} l^{2} \over 2 m r^{2}}
\! \right] \over T^{2B}} \Theta \! \left[  
\mu \!-\! V_{ext}(r) \!-\! {\hbar^{2} l^{2} \over 2 m r^{2}}
\! \right] 
\end{eqnarray}
and
\begin{eqnarray}
\chi(\phi) & \!=\! & \phi \; l \; ,
\end{eqnarray}
\end{mathletters}
where $\Theta\left[ x \right]$ 
denotes the Heavy-side step-function.
To lowest order, the change in the chemical potential
due to the presence of a vortex can be neglected \cite{svidzinsky}.

The method we use to calculate numerically the 
time-evolution of an initial 
wave-packet describing the incoming atom
is based on Cayley's finite difference representation 
of the Schr\"odinger equation \cite{numrec},

\begin{eqnarray}
e^{iH \Delta t} \simeq {1 - i H \Delta t/2 \over 1 + i H \Delta t/2} \; .
\end{eqnarray}
Here, $\Delta t$ is the time step in the discretized 
Schr\"odinger equation. This algorithm is second-order accurate
in time and unitary, i.e., the norm of the wavefunction
is conserved up to computer accuracy.

\begin{figure}[h]
\psfig{figure=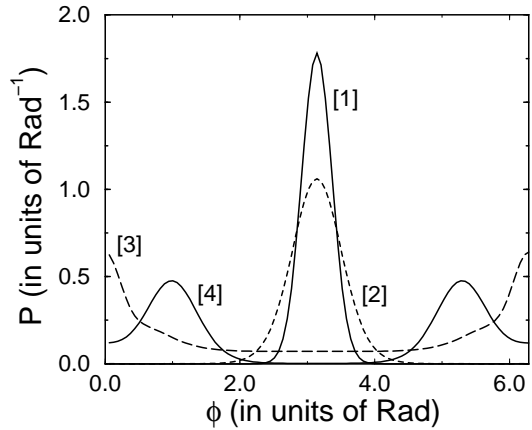}
\vspace{0.25 cm}
\caption{
\label{figmeq0}
\narrowtext 
The angular probability distribution $P(\phi,t)$
at four different times, in the absence of
a vortex. At $t=0$, $P(\phi,t)$ is a Gaussian centered
around $\phi=\pi$, line $\left[1\right]$. Line $\left[4\right]$
is $P(\phi,t)$ after approximately half an
oscillation, $t=3.5 \; \omega_{r}^{-1}$. The short dashed $\left[2\right]$ 
line and the long dashed line 
$\left[3\right]$ are snapshots at
intermediate times, $t=1.8 \; \omega_{r}^{-1}$ and $t=2.0 \; \omega_{r}^{-1}$, 
respectively.
The calculation is performed for $^{85}$Rb in a trap with
$\omega_{r}/2 \pi=200 Hz$.}
\end{figure}

\begin{figure}[h]
\psfig{figure=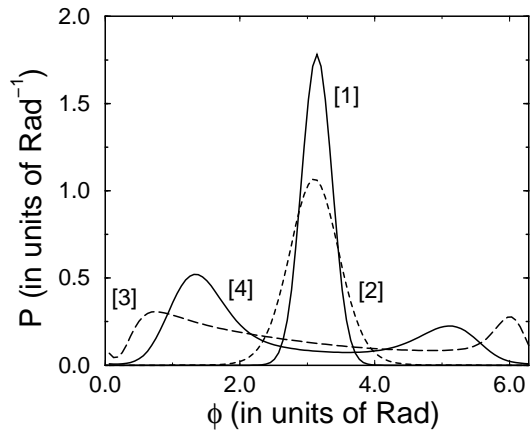}
\vspace{0.25 cm}
\caption{
\label{figmeq1}
\narrowtext 
The angular probability distribution $P(\phi,t)$
at four different times, in the presence of
a vortex with $l=1$. The labeling of the curves,
and the times to which they correspond are
the same as in Fig.~\ref{figmeq0}.}
\end{figure}

The results of our numerical calculations are show in
Fig.~\ref{figmeq0} and Fig.~\ref{figmeq1}, corresponding
to a condensate without and with a vortex, respectively.
In these calculations, the initial wavefunction of the incoming atom is given
by the product of a Gaussian in both the $r$ and the $\phi$
direction, centered around $r=r_{0}$ and $\phi=\pi$.
In principle, we could have plotted the probability
distribution $P(r,\phi,t) \equiv |f(r,\phi,t)|^{2}-|g(r,\phi,t)|^{2}$
as a function of $r$ and $\phi$. 
As time evolves, the hole part $g(r,\phi,t)$ of the atomic wavefunction 
becomes nonzero if the particle part $f(r,\phi,t)$ starts to overlap
with the condensate. Also, the probability distribution
$P(r,\phi,t)$ develops nodes in the radial direction. 
The wavelength of these nodes is a measure of the
kinetic energy of the incoming particle.
However, to present our data more comprehensively, we integrate 
with respect to $r$ and denote the resulting angular probability
distribution by $P(\phi,t)$. Note that $P(\phi,t)$ is normalized to one,
and that this normalization is conserved numerically.
The integrated distribution is shown at four different
times, corresponding to $t=0$, approximately half a period of oscillation
$t=3.5 \; \omega_{r}^{-1}$, and two snapshots at 
$t=1.8 \; \omega_{r}^{-1}$ and $t=2.0 \; \omega_{r}^{-1}$.
 
There are three main differences between 
Fig.~\ref{figmeq0} and Fig.~\ref{figmeq1}.
First, in Fig.~\ref{figmeq0} there is scattering in the
forward direction, whereas in Fig.~\ref{figmeq1} there is
almost no probability for scattering in this direction.  
Second, in Fig.~\ref{figmeq0}, there is almost no scattering in the
backward direction, whereas in Fig.~\ref{figmeq1} there is
scattering in this direction. 
Third, and most important, in Fig.~\ref{figmeq0} the scattering is 
symmetric around $\phi = \pi$, 
whereas in Fig.~\ref{figmeq1} this symmetry is clearly
broken.
The first and second point are related to the
presence of the vortex core, which causes the reflection
in the forward direction to be decreased in favor of the
backward direction.
This has been checked numerically by 
inserting the core, but not the phase of the vortex into
the Bogoliubov-deGennes equations and solving for 
the wavefunction of the scattering particle.
The third point however, cannot be related to the 
presence of the core, which is also rotationally symmetric.
Instead, it is related to the phase of the condensate, which
in the presence of a vortex indeed breaks rotational invariance.
To make this point more clear, we present here a qualitative argument
based on a semi-classical calculation of the Berry-phase associated with
an atom moving around the vortex. The calculation
shows that there is an Aharonov-Bohm effect \cite{aharonov}
in both the particle and the hole parts of the wavefunction, which 
causes the total atomic wavefunction to pick up a non-trivial phase factor.

The Berry-phase $\theta$ associated with the
spin-like nature of a two-component wavefunction
is the phase picked up when the atom 
moves adiabatically along a classical trajectory $C$ \cite{berry}.
If the two-component wavefunction $|\psi({\bf x})\rangle$ describes the
`spin' degrees of freedom, it can be expressed as 

\begin{eqnarray}
\theta & = & i \int_{C} d{\bf x} \cdot \langle \psi({\bf x})|
\nabla |\psi({\bf x})\rangle  \; . 
\end{eqnarray}
In the present case of the Bogoliubov-deGennes equations, we thus get

\begin{eqnarray}
\label{berry}
\theta & = & i \int_{C} d{\bf x} \cdot \left[
f^{*}({\bf x}) \nabla f({\bf x}) - g^{*}({\bf x}) \nabla g({\bf x})
\right] \nonumber \\
& = & - {m \over  \hbar} \int_{C} d{\bf x} \cdot {\bf v}_{s}({\bf x}) 
\left[
|f({\bf x})|^{2} + |g({\bf x})|^{2}
\right] \; .
\end{eqnarray}
Here, the superfluid velocity 
${\bf v}_{s} = \hbar \nabla \chi / 2m$, and 
the coherence factors $f({\bf x})$ and $g({\bf x})$
correspond to one of 
the local eigenfunctions of the Bogoliubov-deGennes equations, i.e., 
without the kinetic term $\hbar^{2} \nabla^{2} / 2 m$.
In the presence of a vortex located at the
origin with winding number $l$, the Berry-phase
for the special case of a circular trajectory at 
radius $r$ therefore becomes

\begin{eqnarray}
\theta & = & - 2 \pi l \left[
|f(r)|^{2} + |g(r)|^{2}
\right]  \; .
\end{eqnarray}
This indeed shows that there is a nontrivial phase accumulated
when the atom moves around the vortex,
which causes the interference pattern to become asymmetric
with respect to $\phi=\pi$. 
Note that it is essential that
the particle and the hole components of the atomic wavefunction contribute with
a different sign to the integral in Eq.~(\ref{berry}).
To understand the direction of the shift in the
interference pattern, 
consider two interfering paths that enclose the vortex.
Seen from $\phi=\pi$, one is 
passing on the left side, and the other is
passing on the right side of the vortex.
Because the accumulated phase for the contour $C=C_{R}-C_{L}$
is negative in the presence of a vortex with $l=1$, we again get constructive
interference if  $C_{L}$ becomes shorter and $C_{R}$ longer. Therefore
the interference pattern for a particle coming from $\phi=\pi$ shifts
towards $\phi=0$, as seen numerically.

In summary, we have presented a possible method 
to detect the presence of a vortex. This is realized by
colliding atoms with the Bose-condensed cloud 
that are identical to the atoms in the condensate. 
The most important signature of the vortex is an
angular asymmetry in the scattered wavefunction, which
we have related to the Aharonov-Bohm phases picked up
by the particle and hole parts of the wavefunction. We note 
that to use the detection mechanism in the way presented here, 
the trap should not rotate. Thus after creating a vortex,
the rotation of the trap must be stopped, in which case the vortex 
in principle becomes unstable. However, recent studies have shown that 
the time needed for the vortex to disappear from the condensate 
diverges with the number of particles in the condensate
at least as $N_{0}^{2/5}$ \cite{feder99,fedichev}. 
Thus, for large $N_{0}$ this presents no difficulty 
for our detection scheme. 
Moreover, a detailed picture of the dynamics of the unstable vortex
might be obtain by means of multiple collisions. 
In this paper however, we considered 
only a time-independent vortex configuration, and the effect
of vortex dynamics on the scattering wavefunction 
is a topic of future investigation.

We are grateful to Eric Cornell for suggesting to us that a
low energetic beam of incoming atoms can be created
by extracting these atoms from the
condensate. We also acknowledge useful
discussions with Eugene Zaremba.

\end{multicols}


\begin{thebibliography}{99}
\bibitem{onsager} L. Onsager, Nuovo Cimento {\bf 6}, 249 (1949).
\bibitem{feynman} R.P. Feynman, in {\it Progress in Low Temperature Physics}
edited by C.J. Gorter (North-Holland, Amsterdam, 1955), Vol. I.
\bibitem{yarmchuk} E.J. Yarmchuk, M.J.V. Gordon, and R.E. Packard,
Phys. Rev. Lett. {\bf 43}, 214 (1979).
\bibitem{packard} R.E. Packard and T.M. Sanders Jr., 
Phys. Rev. Lett. {\bf 22}, 823 (1969); Phys. Rev. A {\bf 6}, 799 (1972).
\bibitem{stringari99} S. Stringari, Phys. Rev. Lett. {\bf 82} 4371 (1999).
\bibitem{baym96} G. Baym and C.J. Pethick, Phys. Rev. Lett. {\bf 76},
6 (1996).
\bibitem{dalfovo96} F. Dalfovo and S. Stringari, Phys. Rev. A {\bf 53},
2477 (1996).
\bibitem{rokhsar97} D.S. Rokhsar, Phys. Rev. Lett. {\bf 79}, 2164 (1997).
\bibitem{fetter} A.L. Fetter, J. Low Temp. Phys. {\bf 113}, 189 (1998).
\bibitem{dodd97} R.J. Dodd, K. Burnett, M. Edwards, and C.W. Clark,
Phys. Rev. A {\bf 56}, 587 (1997).
\bibitem{lundh} E. Lundh, C.J. Pethick, and H. Smith, Phys. Rev. A {\bf 58},
4816 (1998).
\bibitem{bolda} E.L. Bolda and D.F. Walls, Phys. Rev. Lett. {\bf 81}, 5477
(1998).
\bibitem{tempere} J. Tempere and J.T. Devreese, 
Solid State Commun. {\bf 108}, 993 (1998).
\bibitem{sinha} S. Sinha, Phys. Rev. A {\bf 55}, 4325 (1997).
\bibitem{zambelli} F. Zambelli and S. Stringari, Phys. Rev. Lett. {\bf 81},
1754 (1998).
\bibitem{svidzinsky} A.A. Svidzinsky and A.L. Fetter, 
Phys. Rev. A {\bf 58}, 3168 (1998).
\bibitem{goldstein98} E.V. Goldstein, E.M. Wright, and P. Meystre, 
Phys. Rev. A {\bf 58}, 576 (1998).
\bibitem{feder99} D.L. Feder, C.W. Clark, and B.I. Schneider, 
Phys. Rev. Lett. {\bf 82}, 4956 (1999). 
\bibitem{vortexpriv} Very recently the JILA group has succeeded
in creating a vortex-like configuration in an optically trapped
two-component Bose gas (D.S. Hall, private communication).
\bibitem{andreev} A.F. Andreev, Zh. Eksp. Teor. Fiz. {\bf 46}, 182 (1964)
\bibitem{comment}
An even better analogy would be a similar experiment with a Fermi gas
to detect the magnitude of the BCS gap parameter. Unfortunately, it turns
out that in general there is hardly any signature of the BCS
transition in the scattering wavefunction, basically because
in this case the mean-field interactions are too small compared
to the Fermi energy.
\bibitem{campbell98} C.E. Campbell, E. Krotscheck, and M. Saarela,
Phys. Rev. Lett. {\bf 80}, 2169 (1998).
\bibitem{cornellpriv} E.A. Cornell (private communication).
\bibitem{numrec} {\it Numerical Recipes}, W.H. Press, S.A. Teukolsky,
W.T. Vetterling, and B.P. Flannery (Cambridge University, New York, 1992).
\bibitem{aharonov} Y. Aharonov and D. Bohm, Phys. Rev. {\bf 115}, 485 (1959).
\bibitem{berry} M.V. Berry, Proc. R. Soc. A {\bf 392}, 45 (1984).
\bibitem{fedichev} P.O. Fedichev and G.V. Shlyapnikov, 
(unpublished, cond-mat/9902204).
\end{thebibliography}
\end{document}